# Proposed congestion control method for cloud computing environments

Shin-ichi Kuribayashi[1]

[1]Department of Computer and Information Science, Seikei University, Japan
E-mail: kuribayashi@st.seikei.ac.jp

## Abstract

*As cloud computing services rapidly expand their customer base, it has become important to share cloud resources, so as to provide them economically. In cloud computing services, multiple types of resources, such as processing ability, bandwidth and storage, need to be allocated simultaneously. If there is a surge of requests, a competition will arise between these requests for the use of cloud resources. This leads to the disruption of the service and it is necessary to consider a measure to avoid or relieve congestion of cloud computing environments.*

*This paper proposes a new congestion control method for cloud computing environments which reduces the size of required resource for congested resource type instead of restricting all service requests as in the existing networks. Next, this paper proposes the user service specifications for the proposed congestion control method, and clarifies the algorithm to decide the optimal size of required resource to be reduced, based on the load offered to the system. It is demonstrated by simulation evaluations that the proposed method can handle more requests compared with the conventional methods and relieve the congestion. Then, this paper proposes to enhance the proposed method, so as to enable the fair resource allocation among users in congested situation.*

## 1. Introduction

A cloud computing service allows any terminal to access, when and for the duration it requires, the vast and distributed computing resources (processing ability and storage) available on the network without worrying about the particular locations and internal structures of these resources [1]-[4]. It is assumed that necessary resources are taken from a common resource pool. To provide processing ability and storage, it is also necessary to allocate a network bandwidth to access them. This means that multiple types of resources, such as processing ability, bandwidth and storage, need to be allocated, and it is necessary that individual resource types are allocated not independently but simultaneously [5].

As the use of cloud computing services become widespread, it becomes essential for economical service provision to share cloud resources and allocate cloud resources optimally. However, if there is a surge of requests, a competition will arise between these requests for the use of cloud resources. This may lead to the disruption of the service. Therefore, it is necessary to consider a measure to avoid or relieve congestion. In conventional congestion control methods [6]-[16], even when only a specific resource type is congested, the use of all resource types would be restricted. This brings down the efficiency in the use of other resource types, and consequently the serviceability.

Assuming that multiple types of resources are simultaneously allocated to each service, this paper proposes a new congestion control method ("Method A" hereinafter) which reduces the size of required resource for congested resource type, instead of restricting all service requests when a specific types of resource is congested.

Section 2 explains the cloud resource allocation model, assuming that multiple types of resources are simultaneously allocated to each service. Section 3 explain the overview of Method A and clarifies the user service specifications to support Method A. The algorithm to decide the optimal size of required resource to be reduced is also proposed. Section 4 describes simulation evaluations which confirm the effectiveness of Method A. Section 5 proposes to enhance Method A ("Method A-Revised" hereinafter), so as to enable the fair resource allocation among users in congested situation. Section 6 explains the related work. Finally, Section 7 gives the conclusions. This paper is an extension of the study in [17],[18] and [19].

## 2. Cloud resource allocation model

The resource allocation model for a cloud computing environment is such that multiple resources taken from a common resource pool are allocated simultaneously to each request for a certain period [5]. This paper considers two resource types: processing ability and bandwidth, for the preliminary evaluation. It is assumed that the physical facilities for providing cloud computing services are distributed over multiple centers in order to make it easy to increase the number of the facilities when demand increases, to allow load balancing, and to enhance reliability.

The cloud resource allocation model that incorporates these assumptions is illustrated in Fig. 1. Each center has servers (including virtual servers), which provide processing ability, and network devices which provide the bandwidth to access the servers. The maximum size of processing ability and bandwidth at center j (j=1,2,..,k) is assumed to be $C_{maxj}$ and $N_{maxj}$ respectively. When a service request is generated, one optimal center is selected from among k centers, and the processing ability and bandwidth in that center are allocated simultaneously to the request for a certain period.

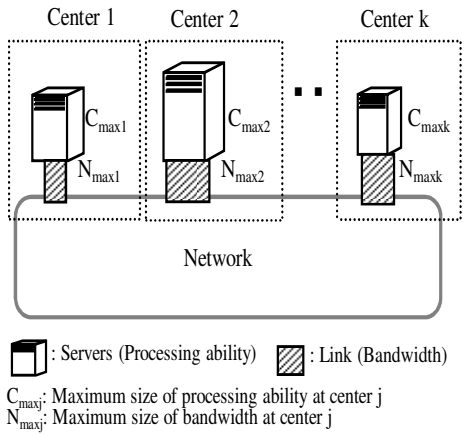

Fig. 1 System model for cloud computing services

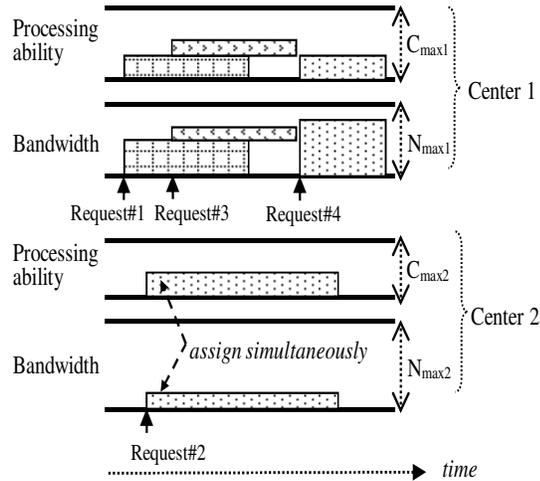

Fig. 2 Example of resource allocation for Fig.1 system model (k=2)

If no center has sufficient resources for a new request, the request is rejected. Fig. 2 illustrates the concept of resource allocation that takes the resource usage period into consideration. These are the same as those in References [5],[18],[19].

# 3. Proposed congestion control method (Method A) for cloud computing environments

## 3.1 Overview of Method A

It is supposed in this paper that there are two types of resources (processing ability and bandwidth), and the size of each resource is normalized based on the maximum resource size for each resource type because units used to specify the sizes of processing ability and bandwidth are different [5].

As discussed in Section 1, the use of all resource types would be restricted even when only a specific resource type is congested in conventional congestion control methods. This brings down the efficiency in the use of other resource types, and consequently the serviceability. The proposed congestion control method, Method A, aims to decrease the request loss probability and to increase the resource efficiency, by reducing the size of required resource for congested resource type, instead of restricting all service requests. For example, the size of required processing ability (congested resource), Cr, will be reduced to its threshold value, Cv, when Cr exceeds Cv. After the size of resources are fixed finally, the joint multiple resource allocation method, Method II in Ref [5], is applied for resource allocation.

Fig. 3 illustrates the image of size reduction of required resource when only processing ability is congested. In this Figure, request type ① requires a large size of processing ability over the threshold Cv , and request type ② does not require a large size of processing ability. As for Request type ①, the system will reduce the size of processing ability. According to the size reduction of processing ability, the size of bandwidth and the value of resource holding time will be also changed. Request type ② will be processed without any reduction of resource size.   Note that details about multiple parameters in Fig. 3 are explained in Section 3.2.

## 3.2 User service specifications

It is proposed to specify the following basic parameters by users to implement Method A as shown in Fig. 4:

<Size of required resource>

·  Size of required processing ability: **$C_r$**

·  Size of required bandwidth: **$N_r$**

<Required resource holding time > **H** (This is the same for two resource types)

In addition to the above parameters, it is proposed to specify the following two new parameters:

<Maximum reduction coefficient> **q** $(0<q\leq1.0)$

This coefficient indicates how much reduction in the size of required resource can be accepted when its resource type is congested. The value of q should be the same for both processing ability and bandwidth. The acceptable minimum resource is $C_0=q*C_r$ for processing ability, and $N_0=q*N_r$ for bandwidth. The length of service time also changes from H into $H_1$ by reducing the size of required resources. The length of $H_1$ will depend on the service the user requests. For example, the service time will be the same for video streaming services even if the video encoding rate is reduced (i.e., the allocated bandwidth is reduced). On the other hand, the service time will become long for file transfer and $H_1$ may be given by H/q.   The other services may require $H_1$ which is more than H/q or less than H/q. Therefore, it was proposed to add one

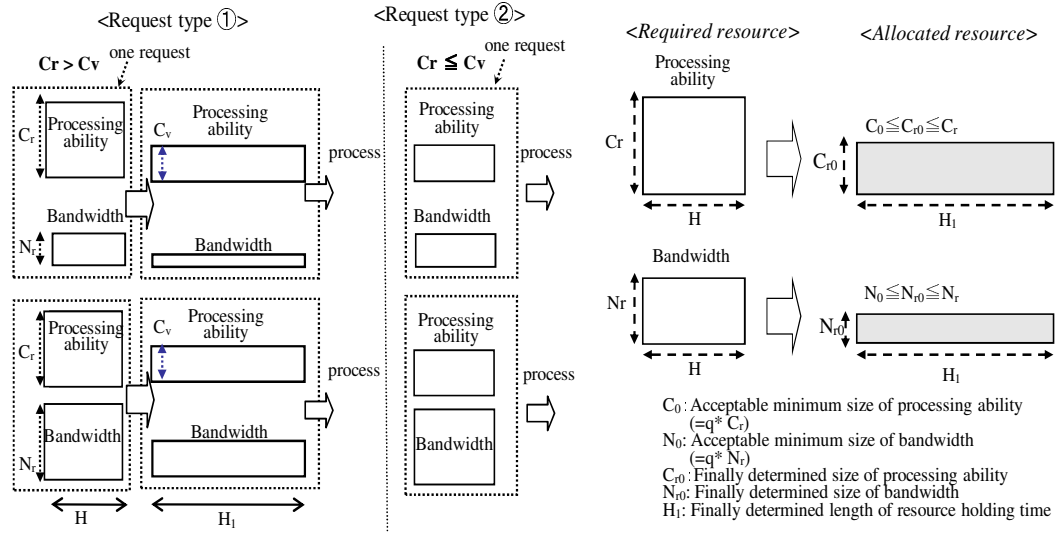

$C_r$: Size of required processing ability, $N_r$: Size of required bandwidth
H: Required resource holding time
$H_1$: Finally determined length of resource holding time
$C_v$: Threshold value for processing ability
$N_c$: Threshold value for bandwidth, M: Ratio of resource holding time
Request type ①: Size of required processing ability is greater than $C_v$
Request type ②: Size of required processing ability is equal to or less than $C_v$

Fig. 3  Image of size reduction of required resource
when processing ability is congested

$C_0$: Acceptable minimum size of processing ability
(=q* $C_r$)
$N_0$: Acceptable minimum size of bandwidth
(=q* $N_r$)
$C_{r0}$: Finally determined size of processing ability
$N_0$: Finally determined size of bandwidth
$H_1$: Finally determined length of resource holding time

Fig. 4 Service specifications for the proposed
congestion control method

additional parameter, Ratio of resource holding time, **M**, and $H_1$ is calculated by

$$H_1 = (h / q) * M \qquad\qquad (1)$$

where a value of M is equal to or greater than that of q.

<Coefficient for the frequency at which the size of required resource is reduced> **p**

Some users or services may not tolerate the size of required resource being reduced too frequently. p ($0 < p \leq 1.0$) is defined for users so as to specify the probability at which their size of required resource is actually reduced in the event of congestion. For example, the size of required resource is always reduced in the event of congestion in the case where p is 1.0, while the size of required resource is reduced in 50% of the cases of congestion (i.e., a reduction in the size is not acceptable in the other 50% of the cases of congestion) in the case where p is 0.5.

## 3.3 Algorithm to decide the optimal size of required resource and reduction timing

This section discusses how much and when the size of required resource is reduced based on the parameters specified by the user, described in Section 3.2. Fig. 4 illustrates an image how to decide the size of required resource when the proposed algorithm is applied. Let $C_{r0}$ ($C_0 \leq C_{r0} \leq C_r$) and $N_{r0}$ ($N_0 \leq N_{r0} \leq N_r$) be the finally determined size of processing ability and bandwidth allocated to a request, and $H_1$ ($H \leq H_1$) the finally determined length of resource holding time. The length of a time block is assumed be much longer than the resource holding time.

## 3.3.1 Algorithm to decide $C_{r0}$, $N_{r0}$, $H_1$

Here, it is supposed that M equals to 1.

(1) The case where only processing ability is congested

Fig. 5 illustrates the procedure flow to decide the values of $C_{r0}$, $N_{r0}$ and $H_1$.

When a new service request is generated, $C_{r0} = C_v$ and $H_1 = (C_r/C_v)*H$ at the probability of p if $C_r$ exceeds the threshold ($C_v$). At the same time, $N_{r0} = (C_v/C_r)*N_r$ at the probability of p. However, if $N_{r0} < N_0$ (i.e., the service requirements are not satisfied), $N_{r0} = N_0$, $C_{r0} = (N_0/N_r)*C_r$ and $H_1 = (N_r/N_0)*H$ are adopted. How to determine the threshold, $C_v$, is discussed in section 3.3.2.

(2) The case where only bandwidth is congested

The very similar procedure flow as in Fig. 5 is applied. If $N_r$ exceeds the threshold value ($N_v$), $N_{r0} = N_v$ and $H_1 = (N_r/N_v)*H$ at the probability of p. At the same time, $C_{r0} = (N_v/N_r)*C_r$ at the probability of p. However, if $C_{r0} < C_0$ (i.e., the service requirements are not satisfied), $C_{r0} = C_0$, $N_{r0} = (C_0/C_r)*N_r$ and $H_1 = (C_r/C_0)*H$ are adopted. How to determine the threshold, $N_v$, is discussed in section 3.3.2.

(3) The case where both processing ability and bandwidth are congested

If $C_r > C_v$, $N_r > N_v$ and $C_v/C_0 > N_v/N_0$, the above procedure (1) is applied. If $C_r > C_v$, $N_r > N_v$ and $C_v/C_0 < N_v/N_0$, the above procedure (2) is applied.

## 3.3.2 Algorithm to decide the optimal values of $C_v$ and $N_v$

As explained in section 3.3.1, the resource subject to size reduction is determined based on the thresholds, $C_v$ and $N_v$. Therefore, the effect of the reduction depends on these thresholds.

(1) Policies to decide the optimal values of $C_v$ and $N_v$

<Policy 1> If lowering $C_v$ or $N_v$ increases the number of requests that can be processed, $C_v$ or $N_v$ is further reduced.

<Policy 2> If lowering $C_v$ or $N_v$ does not increase the number of requests that can be processed, $C_v$ or $N_v$ is kept as high as possible, in order to shorten the time it takes to complete the processing of the service.

(2) The optimal value selection graphs are created in accordance with policies in the above (1) and those are used to determine the optimal values of $C_v$ and $N_v$, based on the load actually offered to the system.   Fig. 6 illustrates one example of the graph for $C_v$.   The vertical axis is the optimal value of $C_v$ and the horizontal axis is the amount of generated service requests. Based on the amount of actually generated service requests, the optimal value of $C_v$ is selected in each time block. The optimal value of $N_v$ is selected in a similar way.

Here, if the load is small enough (Area X in Fig. 6), no congestion occurs and thus there is no need to reduce the size of required resource. Conversely, if the load is too heavy (Area Y in Fig. 6), causing a severe congestion, any change in the size cannot alleviate congestion, and thus it is necessary to implement other congestion control mechanism which restricts most of service requests.

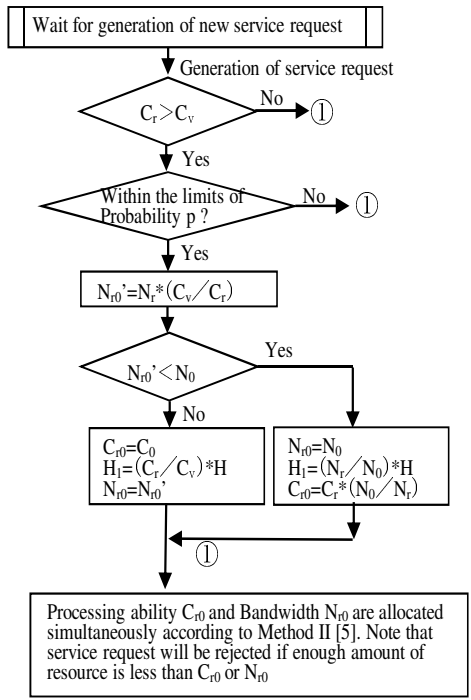

Note: It is supposed here that M is 1.0

Fig. 5 Procedure flow to decide the values of $C_{r0}$, $N_{f0}$ , $H_1$
in the case where only processing ability is congested

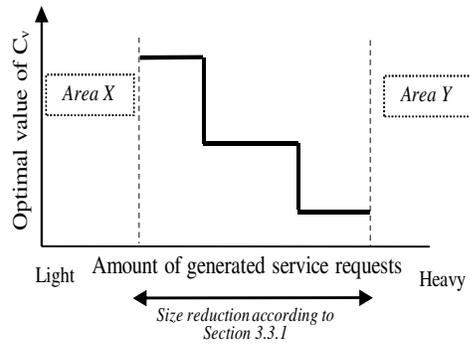

Area X : Congestion control (size reduction) is not applied.
Area Y: Other congestion control mechanism is required.

Fig. 6  Example of graph to decide the optimal value of $C_v$

# 4. Simulation evaluations

## 4.1 Evaluation model

1) The proposed congestion control method, Method A, is evaluated using a simulator written in the C language.

2) Fig. 1 with k=2 is assumed as the resource allocation model. That is, when a new request occurs, one appropriate center between center 1 and center 2 is selected according to the resource allocation algorithm and then both processing ability and bandwidth are allocated simultaneously in the selected center.

3) The size of required processing ability and bandwidth is assumed to follow a Gaussian distribution (its variance was 1). Let C and N be the averages of the distributions of two resource types. The size actually required are respectively $C_r$ and $N_r$ as defined in Section 3.

4) The intervals between requests follow an exponential distribution with the average, r. The length of resource holding time, H, is constant.

5) Ratio of resource holding time M is assumed to be 1.0

6) The pattern in which requests occur is a repetition of {C=$a_1$, N=$b_1$; C=$a_2$, N=$b_2$; …; C=$a_w$, N=$b_w$} , where w is the number of requests that occur within one cycle of repetition, $a_u$ (u=1~w) is the size of C of the u-th request, and $b_u$ (u=1~w) is the size of  N of the u-th request.

## 4.2 Simulation results and evaluation

The simulation results are shown in Figs. 7, 8 and 9. Simulations were conducted assuming that only processing ability was congested in Figs. 7 and 9.   In Fig. 8, it was assumed that both processing ability and bandwidth were congested.   The vertical axis of Fig. 7(1) and Fig. 8, $S_1$, indicated the increased ratio of requests processed after the size of required resource was reduced.   The horizontal axis in Figs. 7 and 8 is the value of $C_v$. Fig. 9 is an example of decision graph of value $C_v$ which is used to decide the optimal value of $C_v$ as explained with Fig. 6 in Section 3.   The following points are clear from these figures:

i) When processing ability is congested, the number of requests that can be processed increases as the value of $C_v$ decreases (Fig. 7(1)). However, the average service completion time is extended when value of $C_v$ decreases (Fig. 7(2)).

［Reason］ When the size of required resource is reduced, the request is likely to be processed even if the amount of available resource is small.

ii) The feature of i) is also applied even when both processing ability and bandwidth are congested (Fig. 8). It is also clear that the value of $S_1$ is large when both processing ability and bandwidth are congested, compared with the case where only processing ability is congested.

[Reasons]   In the case where only processing ability is congested, bandwidth may not be used efficiently though processing ability is congested.   On the other hand, both processing ability and bandwidth will be used efficiently in the case where both processing ability and bandwidth are congested.

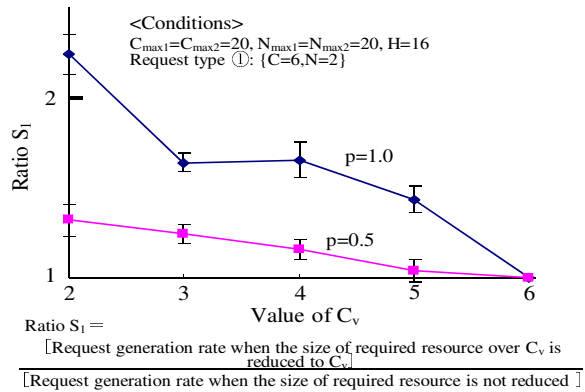

Ratio $S_1$ =

[Request generation rate when the size of required resource over $C_v$ is reduced to $C_v$]
————————————————————————————————————————————
[Request generation rate when the size of required resource is not reduced]

(1) Ratio $S_1$

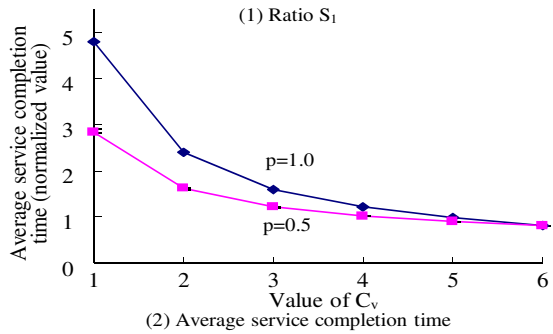

(2) Average service completion time

Fig. 7 Simulation result 1 (the case where only processing ability is congested)

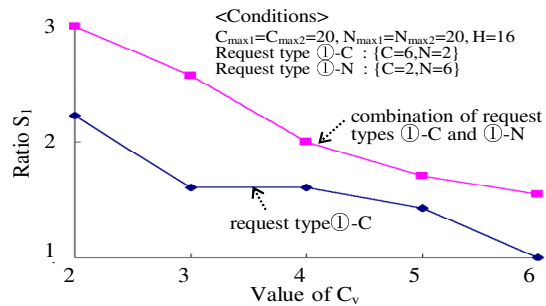

Fig. 8 Simulation result 2 (the case where both processing ability and bandwidth are congested)

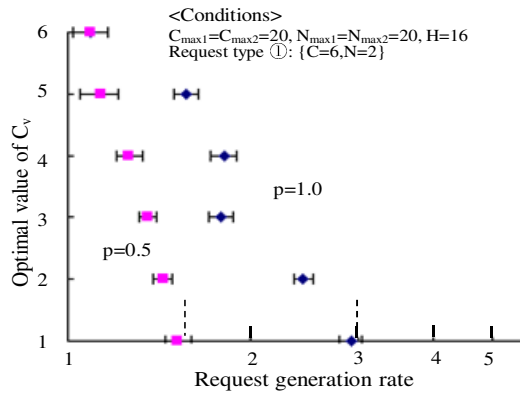

Fig. 9 Decision graph of optimal $C_v$

iii) According to Fig. 9, it is possible to determine the optimal $C_v$ with the actual request generation rate (i.e., the load applied to the cloud computing system). The optimal $C_v$ is 3 in this example when p is 0.5 and the request generation rate is around 1.4. The finally determined size of processing ability and bandwidth allocated to a request ($C_{r0}$, $N_{r0}$) is determined according to the algorithm described in section 3.3.1.

# 5. Fair congestion control algorithm (Method A-Revised)

## 5.1 Unfair resource allocation by Method A

As discussed in Section 4, Method A can achieve an efficient use of resources in congested situation. However, the resources actually allocated to each request can be different depending on their reduction ratios (q) even if two requests require the same resource size. That is, Method A may result in an 'unfair' use of resources.

It is necessary to consider the following two aspects when taking unfairness of Method A into consideration:

(1) Request loss probability

As shown in Fig. 10, the larger the value of q, the larger the required size of resource and consequently, the higher the probability that the request is rejected due to a lack of resource.

(2) Service completion time

The smaller the value of q, the longer the allocated time becomes, and consequently, the later the service completion time becomes.

Since the request loss probability is considered to be more critical than the service completion time in actual services, this paper focuses on the request loss probability in taking unfairness into consideration.

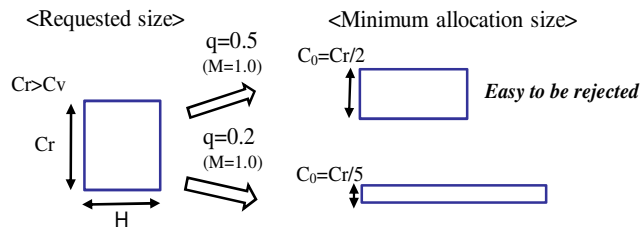

Fig. 10 Example of unfair allocation

## 5.2 Concept of fair resource allocation among users in congested situation

### 5.2.1 Definition of fairness

(1) Fairness can be achieved by putting requests in a queue and allocating resources to them when the required resources become available [20]. However, this paper assumes the loss system (non-delay services) and aims to achieve fairness without queuing in principle.

(2) It is required to consider the fairness assuming that multiple types of resources are allocated to each request simultaneously. If the size of requested resource is relatively large compared to the total resource size, it has a great impact on the process of resource allocation. Therefore, we consider the ratio, W, of the resource size allocated to a request to the maximum resource size of that resource type. We define the resource type that has the largest W as 'key resource type', and propose to focus on the key resource type in considering fairness.

It is noted that the least maximum resource size in all centers is used as for the maximum resource size of a resource type. For example, if the maximum size of processing ability of center 1 is 50, and that of center 2 is 80, 50 is selected as for the maximum resource size of processing ability. Note that the key resource type is not always the same but can change depending on time and users.

(3) As is proposed in references [5], we consider that the fairness can be achieved by allocating key resources to each user in every time block, in proportion to the expected amount of resources requested by each user.

(4) If no user has experienced their requests being rejected, it is determined that no unfairness has occurred during that time block even if the amount of resources allocated to individual users is different.

### 5.2.2 Measure of fair allocation

On the basis of the definition of fairness in Section 5.2.1, we propose to use the following measure for fair resource allocation among users. The number of users is expressed as G in the following:

(1) We introduce the normalization of resource requirement for fairness evaluation. First, define $R_g$ (g=1~G) as the expected amount of resources requested by user g divided by the maximum size of that resource type. $R_g$ is calculated per resource type and the largest value is selected as for user g finally.

(2) Next, $r_g$, is defined as ratio of smallest $R_0$ among all users to $R_g$. For example, if $R_g$ of users 1, 2 and 3 are 100, 50 and 75 respectively, $r_1$=1/2, $r_2$=1 and $r_3$=2/3.

According to the definition of fairness in Section 5.2.1, it can be considered that the resource allocation is 'fair' when the allocated resource size multiplied by $r_g$ is the same for all users. Fig. 11 illustrates one example. In this figure, it is supposed that there are two users and $R_1$ is four times larger than $R_2$ (therefore $r_1$ is 1/4 and $r_2$ is 1). Case A in this figure is considered as a fair but case B is unfair.

(3) For each time block, the key resource type of each user is identified, and the relative value of the total amount of allocated key resource divided by the maximum resource size is calculated. Let $V_i(g)$ be the relative value of total amount of allocated key resource of user g in i-th time block multiplied by $r_g$.

(4) Let $g_1$ be the user with the largest $V_i(g)$, and $N_i(g)$ be the difference between $V_i(g_1)$ and $V_i(g)$. $N_i(g)$ is calculated by

$$N_i(g) = V_i(g_1) - V_i(g) \qquad (2)$$

We consider $N_i(g)$ as the 'imbalance' on allocated resources for user g in i-th time block. Note that $N_i(g_1)$ is equal to 0. Fig. 12 illustrates the relation between $V_i(g)$ and $N_i(g)$ .

(5) It is proposed to check the value **F** given by Equation (3) and to judge that the

smaller the value of F is, the fairer the resource allocation is:

$$F = \left[ \sum_{i=1}^{S} \left\{ \sum_{g=1}^{G} N_i(g) \right\} \right] / S \qquad (3)$$

where S is the total number of time blocks.   Fig. 13 illustrates the meaning of formula F.

   If the value of F is the same for multiple users, it is proposed to judge that the smaller change of the value of $N_i(g)$ is, the fairer the resource allocation is. The change of the value of $N_i(g)$, F1 , can be estimated by

$$F1 = \sum_{i=1}^{S} \left[ \sum_{g=1}^{G} \{ N_i(g) - N_{ave}(g) \}^2 \right] / S \qquad (4)$$

where $N_{ave}(g)$ is the average of $N_i(g)$ in all time blocks.

## 5.3 Fair congestion control algorithm (Method A-revised)

To achieve fair resource allocation in a normal state, the authors proposed an algorithm that attempts to resolve imbalance in resource allocation by allowing the user who has been allocated a smaller resource amount than others to get a provisional resource allocation in the next time section [5]. However, it is difficult to apply the same algorithm to a congested situation as the basic resource allocation mechanism is completely different.

   We propose a new algorithm (Method A-revised) to ensure fairness, which discards requests from the user who had been allocated a relative large resource amount of the key resource type in the previous time block, rather than reducing the resource size allocated to that user.

   Specifically, using $V_i(g)$ calculated by Eq. (5), $\gamma$ % of requests from user g are discarded in time block, i+1, by probability, **$P_i(g)$**, which is calculated by

$$P_i(g) = \frac{V_i(g)}{\displaystyle\sum_{g=1}^{G} V_i(g)} \qquad (5)$$

where $\gamma$ (0<$\gamma \leqq$ 100) is the percentage of the requests that will be rejected to make up for

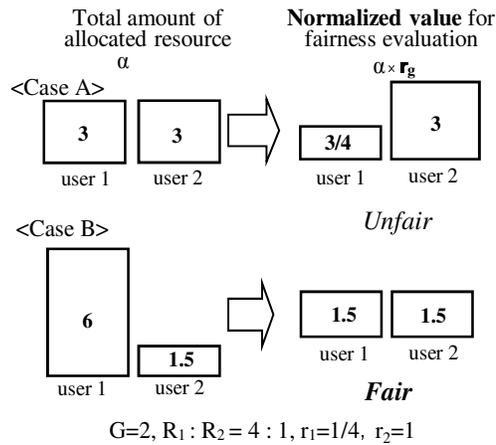

Fig. 11  Normalization of required resource requirement for fairness evaluation

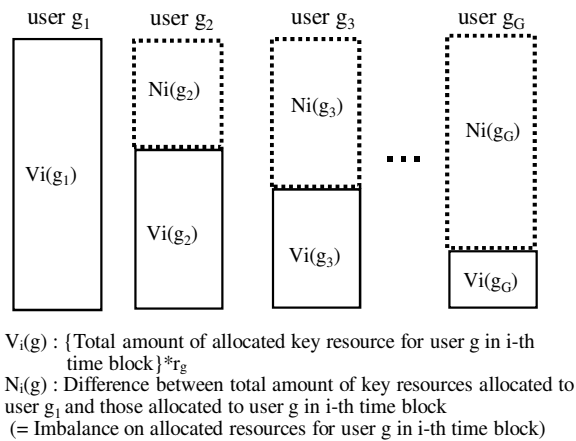

$V_i(g)$ : {Total amount of allocated key resource for user g in i-th time block}*$r_g$

$N_i(g)$ : Difference between total amount of key resources allocated to user $g_1$ and those allocated to user g in i-th time block (= Imbalance on allocated resources for user g in i-th time block) the imbalance in time block. For example, if γ is 80, Method A-revised is applied to allocate resources to 80% of requests from any user but Method A in Section 4 is applied to 20% of requests. The value of γ depends on service requirements. We suppose that the value of γ is common to all users.

Fig. 12  Calculation of $N_i(g)$ in i-th time block

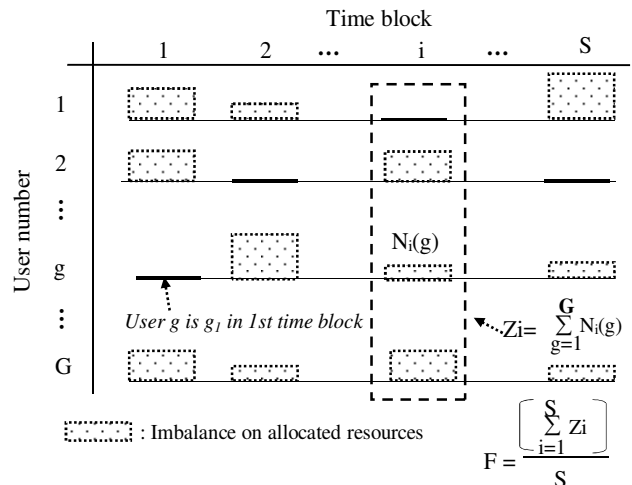

Fig. 13  Meaning of Formula F

## 5.4 Simulation evaluations

### 5.4.1 Conditions

The same conditions in Section 4.1 are applied except for the followings:

1) The expected amount of resources requested by user 1 is the same as that by user 2. That is, $r_1$=1 and $r_2$=1.

2) Ratio of resource allocation time M is assumed to be 1.0 and γ in Section 5.3 is 100.

### 5.4.2 Simulation results and evaluation

The simulation results are illustrated in Fig. 14. This figure illustrates F values (normalized) and the resource efficiencies of Method A and Method A-revised. It is assumed that the average generation interval of requests by user 1 is Y times of user 2 (1≦Y). That is, Y times more requests will be generated from user 2. It is clear that Method A-revised can decrease the value of F greatly (that is, Method A-revised enables fair resource allocation), compared with Method A which does not consider fair allocation.

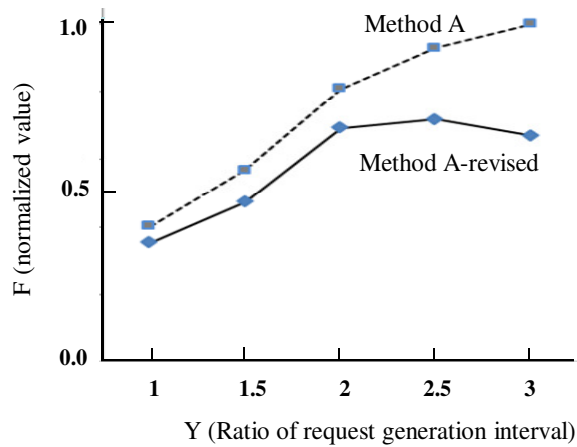

$C_{max1}$=$C_{max2}$=20, $N_{max1}$=$N_{max2}$=20, H=6, q=0.25
{C=2, N=1},$r_1$=1 for user 1; {C=2, N=1}, $r_2$=1 for user 2
Average generation interval of requests by user 1 is Y times of user 2

Fig. 14 Evaluation of fairness (value F)

# 6. Related work

A variety of measures to deal with congestion have been adopted in a variety of existing networks, such as telephone networks, packet-switched networks, mobile phone networks, frame relay networks, ISDN networks, ATM networks, and Internet[6]-[16]. As for Internet, a special effort has been made to study the problems associated with TCP congestion control mechanisms and several solutions that have been proposed to improve its performance [6],[7]. Most of these conventional congestion control methods assume that a single type of resource is allocated to each request and do not cover the model in which both processing ability and bandwidth, dedicated to each request, are rented out simultaneously on a hourly basis. Moreover, the use of all resource types would be restricted in the conventional congestion control methods, even when only a specific resource type is congested. This brings down the efficiency in the use of other resource types, and consequently the serviceability.

The proposed congestion control method (Method A) handles the case where multiple types of resources are allocated simultaneously to each service request and adopts the measure to improve all types of cloud resource. It is also proposed to support the fair resource allocation even in congested situations.

## 7. Conclusions

This paper has proposed a new congestion control method (Method A) for cloud computing environments (in which both processing ability and bandwidth are simultaneously allocated), which reduces the size of required resource for congested resource type instead of restricting all service requests. This paper has proposed the user service specifications for Method A, and clarifies the algorithm to decide the optimal size of required resource to be reduced, based on the load offered to the system. It has been demonstrated by simulation evaluations that Method A can handle more requests compared with the conventional methods and can relieve the congestion.

This paper has also proposed to enhance Method A, so as to enable the fair resource allocation among users in congested situation. A definition of fairness in congested situation and a measure for evaluating fair resource allocation has been clarified and it has been demonstrated by simulation evaluations that Method A-revised enables fair allocation compared with Method A which does not consider the fair allocation.

In the future, we will evaluate the impacts of the number of users, the number of resource types, and the number of centers on the performance of the proposed algorithm.

## Acknowledgment

We would like to thank Mr. Shigehiro Tsumura and Mr. Kenichi Hatakeyama for their help with the simulation.

## References

[1] G.Reese: "Cloud Application Architecture", O'Reilly&Associates, Inc., Apr. 2009.

[2] J.W.Rittinghouse and J.F.Ransone: "Cloud Computing: Imprementation, Management, and Security", CRC Press LLC, Aug. 2009.

[3] P.Mell and T.Grance, "Effectively and securely Using the Cloud Computing Paradigm", NIST, Information Technology Lab., July 2009.

[4] P.Mell and T.Grance："The NIST Definition of Cloud Computing" Version 15, 2009.

[5] S.Kuribayashi, "Optimal Joint Multiple Resource Allocation Method for Cloud Computing Environments", International Journal of Research and Reviews in Computer Science (IJRRCS), Vol.2, No.1, pp.1-8, Feb. 2011

[6] G.Hasegawa and M.Murata, "Survey on Fairness Issues in TCP Congestion Control Mechanisms IEICE Trans. on Commun. Vol.E84-B No.6, pp.1461-1472, June 2001.

[7] H.Oda, H.Hisamatsu and H.Noborio, "Design, Implementation and Evaluation of Congestion Control Mechanism for Video Streaming," International Journal of Computer Networks & Communication (IJCNC), Vol.3, No.3, May 2011

[8] I.A.Qazi, T.Znati and L.H.Andrew, "Congestion Control using Efficient Explicit Feedback," INFOCOM2009.

[9] K.S.Reddy and L.C.Reddy, "A Survey on Congestion Control Mechanisms in High Speed Networks," International Journal of Computer Science and Network Security (IJSNS), Vol.8, No.1, Jan. 2008.

[10] S.Ahmad, A.Mustafa, B.Ahmad,A.Bano, and A.S.Hosam, " Comparative study of Congestion Control Techniques in High Speed Networks," International Journal of Computer Science and Information Security (IJCSIS), Vol.6, No.2, 2009.

[11] R.Jain, "Congestion Control and Traffic Management in ATM networks: Recent advances and A Survey," Computer Networks and ISDN Systems, Vol.28, No.13, pp. 1723-1738, Oct. 1996.

[12] F.M.Holness,"Congestion Control Mechanisms within MPLS Networks", Doctor paper, Queen Mary and Westfield College University of London, Sep. 2000.

[13] M.Welzl, "Network Congestion Control: Managing Internet Traffic," Wiley Series on Communications Networking & Distributed Systems, John Wiley & Sons, Ltd, Sep. 2005.

[14] J.Wang, "Congestion Control in Computer Networks: Theory, Protocols and Applications (Distributed, Cluster and Grid Computing)," Nova Science Pub Inc, Oct. 2010.

[15] B.Raghavan, K.Vishwanath, S.Ramabhadran, K.Yocum, and A.C. Snoeren, "Cloud control with distributed rate limiting," ACM SIGCOMM2007, Aug.2007.

[16] M.Gusat, R.Birke and C.Minkenberg, "Delay-based Cloud Congestion Control," Globecom2009.

[17] K.Hatakeyama and S.Kuribayashi, "Proposed congestion control method for all-IP networks including NGN", The 10th International Conference on Advanced Communication Technology (ICACT2008), 06C-02, pp.1082-1087, Feb. 2008 .

[18] K.Hatakeyama, M.Tanabe and S.Kuribayashi, "Proposed congestion control method reducing the size of required resource for all-IP", Proceeding of 2009 IEEE Pacific Rim Conference on Communications, Computers and Signal Processing (Pacrim2009), Aug. 2009.

[19] T.Tomita and S.Kuribayashi, "Congestion control method with fair resource allocation for cloud computing environments," Proceeding of 2011 IEEE Pacific Rim Conference on Communications, Computers and Signal Processing (Pacrim11), pp.1-6, Aug. 2011.

[20] M.Shreedhar and G.Varghese, "Efficient fair queuing using deficit round robin," IEEE/ACM Transactions on Networking, vol.4, No.3, June 1996.

## Author

**Shin-ichi Kuribayashi**    received the B.E., M.E., and D.E. degrees from Tohoku University, Japan, in 1978, 1980, and 1988 respectively. He joined NTT Electrical Communications Labs in 1980.   He has been engaged in the design and development of DDX and ISDN packet switching, ATM, PHS, and IMT 2000 and IP-VPN systems.   He researched distributed communication systems at Stanford University from December 1988 through December 1989. He participated in international standardization on ATM signaling and IMT2000 signaling protocols at ITU-T SG11 from 1990 through 2000.   Since April 2004, he has been a Professor in the Department of Computer and Information Science, Faculty of Science and Technology, Seikei University.   He is a member of IEEE, IEICE and IPSJ.

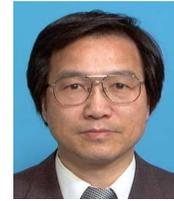